\documentstyle[latexsym,amssymb,aps,floats,eqsecnum,
preprint,amsfonts]{revtex}
\def\laq{\raise 0.4ex\hbox{$<$}\kern -0.8em\lower 0.62 ex\hbox{$\sim$}}
\def\gaq{\raise 0.4ex\hbox{$>$}\kern -0.7em\lower 0.62 ex\hbox{$\sim$}}
\input epsf
\begin{document}

\bibliographystyle{unsrt}

\title{Scalar normal modes of higher dimensional gravitating  kinks}
\author{Massimo Giovannini
\footnote{Electronic address: 
Massimo.Giovannini@ipt.unil.ch}}

\address{{\it Institute of Theoretical Physics, 
University of Lausanne}}
\address{{\it BSP-1015 Dorigny, Lausanne, Switzerland}}

\maketitle

\begin{abstract} 
The scalar normal modes of higher dimensional gravitating 
kink solutions are derived. By perturbing to second order 
the gravity and  matter parts of the action in the 
background of a five-dimensional kink, the effective 
Lagrangian of the scalar fluctuations is derived and diagonalized 
in terms of a single degree of freedom which invariant 
under infinitesimal diffeomorphisms.
The spectrum of the normal modes 
is discussed and applied to the analysis of short distance corrections 
to Newton law.
\end{abstract}
\vskip0.5pc
\centerline{Preprint Number: IPT-UNIl-02-8 }
\vskip0.5pc
\noindent

\newpage
\renewcommand{\theequation}{1.\arabic{equation}}
\setcounter{equation}{0}
\section{Introduction}
Higher dimensional kink solutions \cite{1} 
have been introduced in order to discuss the localization 
properties of fields in the context of infinite 
extra dimensions \cite{2,ak,vis,2a}. Of particular 
interest is, in this framework,  the problem of localization of 
gravitational interactions \cite{3,3a} and of gauge 
interactions \cite{3b,rubr,dv,od,rub,ti,gi}.
 
It has been recently shown that five-dimensional 
gravitating kinks possess and interesting structure 
of zero modes \cite{5} (see also \cite{6a} and \cite{3b}).
In their simplest realization, gravitating 
kinks can appear in five-dimensional 
scalar-tensor theories of gravity when the  five-dimensional 
bulk coordinate is infinite \cite{7,8,8a,9}. The gravity theory can be 
taken to be, for simplicity, of Einstein-Hilbert type 
(appropriately extended to higher dimensions). However,
also quadratic gravity theories with Gauss-Bonnet self-interactions
 allow the same type of static solutions \cite{10,11}. More 
generally, string inspired 
solutions with Gauss-Bonnet terms have been  discussed from 
different points of view \cite{11a,11b,110,12,13,14}.

Gravitating kinks have scalar, vector and tensor normal modes \cite{5}
with respect to four-dimensional
 Poincar\'e transformations which are always unbroken.
The tensor normal modes have been extensively analyzed 
in the context of brane solutions 
with ${\rm AdS}_{5}$ geometries \cite{3,3a} where it has been 
shown that Poincar\'e invariance in four-dimensions implies 
the existence of a localized tensor zero mode. The tower 
of massive states, when resummed, leads to computable 
corrections to the Newton potential at short distances. 
The vector modes of the geometry (the so-called graviphoton fields) 
are not localized on five-dimensional gravitating kinks and 
they are always massless \cite{5,15}.

A different situation appears in the case of scalar degrees of freedom
since they do have continuum modes.
As it will be shown in detail, the scalar fluctuations of the 
action perturbed to second order combine in a single 
 degree of freedom. This combination  diagonalizes the full (second order)
action and it is invariant under infinitesimal diffeomorphisms. 

It is the purpose of the present paper 
to derive precisely the effective Lagrangian of the scalar normal modes of 
the gravitating five-dimensional kink solutions. There is no 
a priori reason to expect the simplicity of the result from the complicated 
structure of the scalar action perturbed to second order. The 
strategy is, in short, the following. Consider the five-dimensional extension of the 
Einstein-Hilbert action minimally coupled 
to a scalar field $\varphi$ :
\begin{equation}
S= \int d^5 x \sqrt{|G|} \biggl[ - \frac{ R}{2 \kappa} + \frac{1}{2} 
G^{A B} \partial_{A} \varphi \partial_{B} \varphi - V(\varphi) \biggr].
\label{action}
\end{equation}
In this framework, kink solutions can be obtained in a 
geometry of the type 
\begin{equation}
ds^2 = a^2(w) [ \eta_{\mu\nu} dx^{\mu} dx^{\nu} - dw^2],
\label{metric}
\end{equation}
where $w$ is the bulk coordinate and $\eta_{\mu\nu}$ is the Minkowski metric. 
The potential appearing in Eq. 
(\ref{action}) is symmetric for $\varphi \to - \varphi$. 
The gravity and matter parts of the action (\ref{action}) will then be 
perturbed to second order in the amplitude of the scalar fluctuations 
of the metric without fixing a specific gauge \footnote{In the present 
paper, the {\em first} order fluctuations of a given quantity will be denoted 
by $\delta^{(1)}$ while the {\em second} order fluctuations will be denoted by 
$\delta^{(2)}$.}:
\begin{equation}
\delta^{(2)} S = \delta^{(2)} S_{\rm gr} + \delta^{(2)} S_{\rm m}. 
\end{equation}
In this procedure various total derivatives appear. Some of them
are expected. For instance the known total derivative coming from the surface 
term of the Einstein-Hilbert action. Other total derivatives 
are accidental in the sense that they appear only when 
the perturbed gravity and matter part of the action are combined 
and evaluated on-shell, i.e. on the  background configuration.

A naive way of thinking would suggest that the perturbation 
of the action to second order should led exactly to the 
same results one would obtain by perturbing to first order in the amplitude 
of the scalar fluctuations 
the equations of motion derived from the action (\ref{action}). This 
procedure has been already discussed and 
the present analysis shows that the naive 
expectation is only partially true.  
From the equations of motion various gauge-invariant quantities 
can be defined all leading to decoupled  equations 
for the fluctuations. These variables, even if invariant 
under infinitesimal diffeomorphisms, do not 
diagonalize the action and, hence, are not the correct normal 
modes of the scalar tensor action (\ref{action}) in the 
background of a gravitating kink. 

The canonical structure of the action is 
particularly important if the fluctuations of a given spin are quantized. It 
would not be correct to quantize a fluctuation whose action is not canonical.
For instance, it could be easily shown that the scalar mode associated 
with the 
fluctuation of the field $\varphi$ obeys a Schr\"odinger-like equation when 
the coupling to the metric fluctuations is ignored. The canonical 
normal modes are the correct classical quantities to be 
promoted to field operators. 

The present paper is organized as follows. In Section II the basic 
properties of the formalism will be introduced. In Section III 
the second order fluctuation of the gravity and matter parts
 of the action will be derived.  
Section IV contains the diagonalization of the full action in terms of its 
canonical normal modes. In Section V the corrections to Newton potential 
coming from the tower of scalar fluctuations of a gravitating kink will be 
addressed. Finally Section VI contains a summary of the main findings and 
some concluding remarks.
For purposes of presentation, various technical details 
needed for the derivations have been collected in the Appendix.

\renewcommand{\theequation}{2.\arabic{equation}}
\setcounter{equation}{0}
\section{Gravitating kinks and their fluctuations}
In the metric (\ref{metric}) the background equations of motion derived 
from the action (\ref{action}) become 
\begin{eqnarray}
&& {\cal H}' + {\cal H}^2 = - 
\frac{1}{6} \biggl[ \frac{{\varphi '}^2}{2} + V a^2 \biggr],
\label{b1}\\
&& {\cal H}^2 = \frac{1}{12}\biggl[ \frac{{\varphi '}^2}{2} - V a^2 \biggr],
\label{b2}\\
&& \varphi'' + 3 {\cal H}\varphi' - a^2 \frac{\partial V}{\partial\varphi} =0,
\label{b3}
\end{eqnarray}
where ${\cal H} = a'/a$ and the prime denotes derivation with respect 
to the bulk coordinate. Notice that natural gravitational units 
$ 2 \kappa =1 $ are used.
Eqs. (\ref{b1})--(\ref{b3}) admit gravitating kink solutions. For instance, 
we will be 
interested in solutions of the type 
\begin{eqnarray}
&& \varphi(w)= \sqrt{6}  \arctan{(b w)},
\label{s1}\\
&& a(w) = ( b^{2} w^{2} + 1)^{ - \frac{1}{2}},
\label{s2}\\
\end{eqnarray}
arising in sine-Gordon potentials 
\begin{equation}
 V(\varphi) = 3 b^2 \biggl[ 5 \cos^2{(\varphi/\sqrt{6})} -4 \biggr],
\label{s3}
\end{equation} 
or in other classes of symmetric potentials 
\cite{7,7a,8,8a,9,10,11,11a,11b,110,12}.

In the case of scalar fluctuations the total metric can be written, 
in its most general form, as 
\begin{equation}
ds^2 = a^2(w)\biggl\{\biggl[\eta_{\mu\nu} + 2 \biggl( \eta_{\mu\nu} 
+ \partial_{\mu\nu} E\biggr)\biggr\}dx^{\mu} dx^{\nu}
+ 2 \partial_{\alpha}C dx^{\alpha} dw -( 1 - 2 \xi)dw^2 \biggr\}
\label{pertme}
\end{equation}
where $\psi$, $\xi$, $C$ and $E$ are four functions 
characterizing the scalar (Poincar\'e-invariant ) 
modes of the first order fluctuations 
of the metric $G_{AB}$, i.e., in our notations, $\delta^{(1)} G_{A B}$.
  
The infinitesimal diffeomorphisms preserving the scalar nature of the 
fluctuation are 
\begin{eqnarray}
&& x^{\mu} \to \tilde{x}^{\mu} = x^{\mu} + \partial^{\mu} \epsilon,
\nonumber\\
&& w \to \tilde{w} = w + \epsilon^{w}.
\label{trans}
\end{eqnarray}
Under the infinitesimal coordinate transformation (\ref{trans}) 
the scalar fluctuations of the metric change as 
\begin{eqnarray}
&&\tilde{E} = E - \epsilon,
\label{El}\\
&&\tilde{\psi} = \psi - {\cal H} \epsilon_{w},
\label{psil}\\
&& \tilde{C} = C - \epsilon' + \epsilon_{w},
\label{Cl}\\
&& \tilde{\xi} = \xi + {\cal H} \epsilon_{w} + \epsilon_{w}'.
\label{xil}
\end{eqnarray}
In spite of this, two gauge-invariant fluctuations can be 
constructed \cite{5}:
\begin{eqnarray}
&&\tilde{\Psi} = \tilde{\psi} - {\cal H}  ( \tilde{E}' - \tilde{C}), 
\label{giscal0}\\
&& \tilde{\Xi} = \tilde{\xi} - \frac{1}{a} [ a( \tilde{C} - \tilde{E}')]'.
\label{giscal}
\end{eqnarray}
The fluctuations of the domain-wall itself (i.e. the fluctuations 
of $\varphi$) 
\begin{equation}
\varphi(x^{\mu}, w) = \varphi(w) + \chi(x^{\mu}, w),~~~~~~
\delta^{(1)} \varphi = \chi,
\label{chifl}
\end{equation}
are also non gauge-invariant
\begin{equation}
\tilde{\chi} = \chi - \varphi'\epsilon_{w}.
\label{gvchi}
\end{equation}
The gauge-invariant scalar field fluctuation will be 
\begin{equation}
\tilde{X} = \tilde{\chi} - \varphi' (\tilde{E}' -\tilde{C}).
\label{chigi}
\end{equation}
It is worth noticing that Eqs. (\ref{giscal0}) and (\ref{giscal})
are reminiscent of the Bardeen potentials which are 
normally introduced in the gauge-invariant theory of gravitational
fluctuations in four-dimensional cosmological backgrounds \cite{bar}. Of 
course the problem treated here is very different: we deal with static 
backgrounds, we are in five dimensions and four-dimensional 
Poincar\'e symmetry (unlike five-dimensional Poincar\'e symmetry) 
is unbroken.

Already from this analysis we can guess that there are 
three gauge-invariant scalar functions subjected 
to the Hamiltonian and momentum constraint. Hence 
only one  degree of freedom describing
the scalar fluctuations of the gravitating kink should be expected. 
This degree of freedom will emerge naturally from the 
structure of the effective action perturbed 
to second order in the amplitude of the 
fluctuations appearing in (\ref{pertme}).

\renewcommand{\theequation}{3.\arabic{equation}}
\setcounter{equation}{0}
\section{Gravity and matter actions to second order}

The gravity part of the action
(\ref{action}) will now be perturbed to second 
order in the amplitude of the scalar fluctuations of the metric.
First of all it should be noticed that the 
 Einstein-Hilbert action can be written in a form 
where the known surface term is already absent, namely, 
\begin{equation}
S_{\rm gr} = -  \int d^{5} x \sqrt{|G|} R = 
 \int d^{5} x \sqrt{|G|}  G^{ A B } \biggl[ 
\Gamma_{ A B}^{M} \Gamma_{M N}^{N} - \Gamma_{ A N}^{M} \Gamma_{M B}^{N}
\biggr].
\end{equation}
Hence the second order fluctuation of the gravity part of the action can be 
written as 
\begin{eqnarray}
\delta^{(2)} S_{\rm gr} &=& 
 \int d^{5} x\biggl\{ \sqrt{|\overline{G}|} \biggl[\delta^{(2)} G^{ A B } \biggl(
\overline{\Gamma}_{ A B}^{M} \overline{\Gamma}_{M N}^{N} - 
\overline{\Gamma}_{ A N}^{M} \overline{\Gamma}_{M B}^{N}
\biggr) 
\nonumber\\
&+& \overline{G}^{ A B } \biggl( 
\delta^{(2)} 
\Gamma_{ A B}^{M} \overline{\Gamma}_{M N}^{N}+ 
\overline{\Gamma}_{ A B}^{M} \delta^{(2)}\Gamma_{M N}^{N} 
- \delta^{(2)} \Gamma_{ A N}^{M} \overline{\Gamma}_{M B}^{N}
- \overline{\Gamma}_{ A N}^{M} \delta^{(2)}\Gamma_{M B}^{N}\biggr)
\nonumber\\
&+&  \overline{G}^{ A B } \biggl( 
\delta^{(1)} 
\Gamma_{ A B}^{M} \delta^{(1)}\Gamma_{M N}^{N}
- \delta^{(1)} \Gamma_{ A N}^{M} \delta^{(1)}\Gamma_{M B}^{N}\biggr)
\nonumber\\
&+& \delta^{(1)} G ^{ A B } \biggl( 
\delta^{(1)} \Gamma_{ A B}^{M} \overline{\Gamma}_{M N}^{N} 
+ \overline{\Gamma}_{ A B}^{M} \delta^{(1)} \Gamma_{M N}^{N}
 - \delta^{(1)}\Gamma_{ A N}^{M} \overline{\Gamma}_{M B}^{N}
- \overline{\Gamma}_{ A N}^{M} \delta^{(1)}\Gamma_{M B}^{N}
\biggr)
\biggr]
\nonumber\\
&+& \delta^{(2)} \sqrt{|G|} \biggl[  \overline{G}^{ A B } \biggl(
\overline{\Gamma}_{ A B}^{M} \overline{\Gamma}_{M N}^{N} - 
\overline{\Gamma}_{ A N}^{M} \overline{\Gamma}_{M B}^{N}\biggr)\biggr]
\nonumber\\
&+&  \delta^{(1)} \sqrt{|G|} \biggl[  \delta^{(1)} G^{ A B } \biggl(
\overline{\Gamma}_{ A B}^{M} \overline{\Gamma}_{M N}^{N} - 
\overline{\Gamma}_{ A N}^{M} \overline{\Gamma}_{M B}^{N}\biggr) +\overline{G}^{ A B } \biggl(
\delta^{(1)}\Gamma_{ A B}^{M} \overline{\Gamma}_{M N}^{N} +
\overline{\Gamma}_{ A B}^{M} \delta^{(1)}\Gamma_{M N}^{N}
\nonumber\\
&-&  
\delta^{(1)}\Gamma_{ A N}^{M} \overline{\Gamma}_{M B}^{N}-
\overline{\Gamma}_{ A N}^{M} \delta^{(1)}\Gamma_{M B}^{N}\biggr)\biggr]
\biggr\},
\label{acsec}
\end{eqnarray}
where $\overline{G}^{A B} $ and $\overline{\Gamma}_{A B }^{C}$ are, 
respectively, 
the background values of the metric and of the Christoffel connections.
In (\ref{acsec}) there are different kinds of contributions coming both from 
the second order fluctuations of the inverse metric (and of its determinant) 
and 
from the second order fluctuations of the Christoffel connections.
All the  results needed in order to obtain the explicit 
form of (\ref{acsec}0 in terms of the degrees of freedom appearing 
in Eq. (\ref{pertme}) are separately reported in the Appendix. 
Thus, using the results of the Appendix, and, in particular, 
inserting Eqs. (\ref{m1})--(\ref{d1}), (\ref{m2})--(\ref{d2}) and 
(\ref{c1})--(\ref{c2}) 
into Eq. (\ref{acsec}) the second order form of the gravity part of 
the action is obtained:
\begin{eqnarray}
\delta^{(2)} S_{\rm gr} &=& \int d^5 x \biggl\{ a^3 \biggl[ {\cal H}^2 
\biggl( 48 \psi^2 + 18 \xi^2 + 
12( \xi + 2 \psi) \Box E + 48 \psi \xi + 6 (\Box E)^2 
\nonumber\\
&-& 6 \partial_{\alpha}C \partial^{\alpha} C
-12 \partial_{\alpha} \partial_{\beta} E \partial^{\alpha} \partial^{\beta} 
E \biggr)
\nonumber\\
&+& {\cal H} \biggl(48 \psi \psi' + 12 \psi \Box E' + 12 \psi' \Box E + 
6 \Box E \Box E' -12 \partial_{\alpha} \partial_{\beta} E' \partial^{\alpha} 
\partial^{\beta} E 
\nonumber\\
&-& 12 
\partial^{\alpha} C \partial_{\alpha} \psi + 3 \partial^{\alpha } 
C \partial_{\alpha} \xi 
+ 24 \xi \psi' + 6 \xi \Box E'\biggr)
\nonumber\\
&-& 6 \partial_{\alpha} \psi ( \partial^{\alpha} \psi - \partial^{\alpha} \xi)
+ 12 {\psi'}^2 + 6 \psi' \Box E' - 2 \psi' \Box C - 4 \partial_{\alpha} C' 
\partial^{\alpha} \psi 
\nonumber\\
&+& \xi' \Box C - \partial^{\alpha} C' \partial_{\alpha } \xi \biggr] 
+ {\cal D}_{1} + {\cal D}_2
+ {\cal D}_{3} 
\biggr\},
\label{acgrav}
\end{eqnarray}
 where ${\cal D}_1$, ${\cal D}_{2}$ and ${\cal D}_{3}$ are the following total 
derivatives 
\footnote{Notice that, in order to shorten the notation, the convention 
$\Box = \eta^{\alpha\beta} \partial_{\alpha} 
\partial_{\beta}$ has been adopted.}
\begin{eqnarray}
{\cal D}_{1} &=& \partial_{\alpha} \biggl\{ - a^3 {\cal H} \biggl[ 
3 \partial^{\alpha} C ( 4 \psi + 
\Box E ) + 3 \xi \partial^{\alpha} C + 6 \partial^{\alpha} C \psi + 
\partial^{\alpha} \partial^{\beta} E \partial_{\beta} C\biggr] \biggr\},
\nonumber\\
{\cal D}_{2} &=& \partial_{\alpha} \biggl\{ a^3 \biggl[ \partial^{\alpha} 
\partial_{\beta} E 
\partial^{\beta} \Box E - \partial^{\nu} \partial^{\beta} E \partial^{\alpha} 
\partial_{\beta} \partial_{\nu} E
+ 2 \partial_{\beta} C \partial^{\alpha} \partial^{\beta} E'
\nonumber\\
&-& 2 \partial^{\alpha} C \Box E' - 2\partial^{\alpha} C' \Box E 
+ \partial^{\alpha} E' \Box E' - \partial_{\beta} E' \partial^{\alpha} 
\partial^{\beta} E'\biggr],
\nonumber\\
{\cal D}_{3} &=& \biggl\{ a^3 \Box C \Box E \biggr\}'.
\end{eqnarray}

The same procedure discussed in the case of the  gravity part 
of the action, should be repeated for the matter part. 
recalling the notation for the fluctuations of the domain-wall 
itself, i. e. Eq. (\ref{chifl}), 
the perturbed matter part of the action can be written, in general terms,
as
\begin{eqnarray}
\delta^{(2)} S_{\rm m} &=& \int d^5 x \frac{1}{2} \biggl\{
\delta^{(2)} \sqrt{ |G|}\biggl[
\overline{G}^{A B} \partial_{A} \varphi \partial_{B} \varphi - V(\varphi)
\biggr]
\nonumber\\
&+&  \sqrt{ | \overline{G}|}\biggl[ \delta^{(2)} 
G^{A B} \partial_{A} \varphi \partial_{B} \varphi + \overline{G}^{ A B} 
\partial_{A } \chi \partial_{B} \chi 
\nonumber\\
&+& \delta^{(1)} 
G^{A B}( \partial_{A} \varphi \partial_{B} \chi 
+\partial_{A} \chi \partial_{B} \varphi )
 -\frac{1}{2} \frac{\partial^2 V}{\partial\varphi^2} \chi^2
\biggr]
\nonumber\\
&+& \delta^{(1)} \sqrt{ |G|}\biggl[
\delta^{(1)} G^{A B} \partial_{A} \varphi \partial_{B} \varphi+ 
 \overline{G}^{A B} ( \partial_{A} \chi \partial_{B} \varphi + 
\partial_{A} \varphi \partial_{B} \chi)
 - \frac{\partial V}{\partial\varphi} \chi \biggr] \biggr\}. 
\label{masec}
\end{eqnarray}
Using now Eqs. (\ref{m1})--(\ref{d1}) and Eqs. (\ref{m2})--(\ref{d2})
into Eq. (\ref{masec}) we get the explicit form of the 
matter action perturbed to second order:
\begin{eqnarray}
\delta^{(2)} S_{\rm m} 
&=& \int d^5 x \biggl\{ \frac{1}{2}\biggl( \frac{{\varphi}'}{2} 
+ V a^2\biggr) \biggl[ \xi^2
- 8 \psi^2 - (\Box E)^2 + 2\partial_{\alpha} \partial_{\beta} 
E \partial^{\alpha} 
\partial^{\beta} E - 4 \psi \Box E 
\nonumber\\
&+& 8 \psi \xi + 2 \xi \Box E - \partial_{\alpha} 
C \partial^{\alpha} C\biggr]
\nonumber\\
&+& ( \xi - 4 \psi - \Box E) \biggl[ \chi' \varphi' + {\varphi'}^2 \xi 
+ a^2 \frac{\partial V}{\partial\varphi} \chi\biggr] 
-\frac{1}{2} {\chi'}^2 + \partial_{\alpha} \chi \partial^{\alpha} \chi + 
\varphi' \partial_{\alpha} \chi \partial^{\alpha} C 
\nonumber\\
&-& 2 \varphi' \chi' \xi + \frac{{\varphi'}^2}{2} 
\biggl[ \partial_{\alpha} C\partial^{\alpha} C - 4 \xi^2 \biggr] -   
\frac{1}{2}\frac{\partial^2 V}{\partial\varphi^2}\chi^2\biggr\}.
\label{acmatt}
\end{eqnarray}
The contributions to the second order action coming, respectively,
 from the gravity (\ref{acgrav}) and matter parts (\ref{acmatt}) should 
be combined. In this procedure  various simplifications occur. 
First of all, since the action should be evaluated on the background of 
the gravitating kink, Eqs. 
(\ref{b1})--(\ref{b3}) can be imposed. 
The resulting  (total) action is, therefore, 
\begin{eqnarray}
\delta^{(2)} S &=& \delta^{(2)}S_{\rm gr} + \delta^{(2)} S_{\rm m}
\nonumber\\
&=& \int d^{5} x  \biggl\{ a^3 \biggl[12 {\psi'}^2 + 
3( {\cal H}' + 3 {\cal H}^2) \xi^2 - 6 \partial_{\alpha} \psi ( 
\partial^{\alpha} \psi - \partial^{\alpha} \xi) + 24 {\cal H} \xi \psi'
\nonumber\\
&+& (\xi' + 4 \psi ') \varphi' \chi + 2 \xi \frac{\partial V}{\partial\varphi} 
a^2 \chi - \frac{1}{2} {\chi'}^2 + \frac{1}{2} \partial_{\alpha} \chi 
\partial^{\alpha} \chi
\nonumber\\
&-& \frac{1}{2} \frac{\partial^{2} V}{\partial\varphi^2}
a^2 \chi^2 + \Box( E' - C) ( 6 {\cal H} \xi + 6 \psi' + \varphi' \chi)
\biggr] 
\nonumber\\
&+& {\cal D}_{1} + {\cal D}_{2} + {\cal D}_{3}+
{\cal D}_{4} + {\cal D}_{5} \biggr\}, 
\label{consac}
\end{eqnarray}
where ${\cal D}_{4}$ and  ${\cal D}_{5}$  are two further 
 total derivatives which appear as a consequence 
of the use of the background equations (\ref{b1})--(\ref{b3}) 
in the perturbed action
\begin{eqnarray}
{\cal D}_{4} &=& \partial_{\alpha} \biggl\{ a^3\biggl[ 12 {\cal H} \biggl( 
2\partial^{\alpha} E' \Box E-  2\partial_{\beta} E' 
\partial^{\alpha}\partial^{\beta} E\biggr) + 6({\cal H}' + 3 {\cal H}^2)
\biggl( \partial^{\alpha} E \Box E - \partial_{\beta} E \partial^{\alpha}
 \partial^{\beta} E\biggr)
\nonumber\\
&+& {\cal H}\biggl( (12 \psi - 3 \xi+ \varphi' \chi) 
\partial^{\alpha} C  - ( 4 \psi + \xi) \partial^{\alpha} C'\biggr)\biggr]
\biggr\},
\nonumber\\
{\cal D}_{5} &=& \biggl\{ a^3 \biggl[ - 3 {\cal H} (\Box E)^2 
+ 12 {\cal H} \psi \Box E + 24 {\cal H} \psi^2 - \varphi' \chi\Box E + 
(4 \psi + \xi) \Box C 
\nonumber\\
&-& ( \xi + 4 \psi ) \varphi' \chi\biggr]\biggr\}'.
\end{eqnarray}
The action obtained in Eq. (\ref{consac}) will now be diagonalized.

\renewcommand{\theequation}{4.\arabic{equation}}
\setcounter{equation}{0}
\section{Canonical normal modes of the second order action}

The variation of the action (\ref{consac}) with respect to $(E'- C)$ leads 
to the constraint
\begin{equation}
6{\cal H} \xi  + 6 \psi' + \varphi' \chi =0.
\label{momcon}
\end{equation}
From the gauge-invariant analysis of the evolution equations 
of the fluctuations we do know that there are 
variables obeying simple (Schr\"odinger-like) equations.
However, we cannot say, only from the equations of motion, that these 
variables diagonalize the second order action. 

In order to diagonalize the action (\ref{consac}) let us look first at a
small portion of it, namely the kinetic terms of the 
various fluctuations. If a variable diagonalizing the 
kinetic terms can be found, then it will be worth trying to see if also 
all the other terms of (\ref{consac}) will be diagonal in the same variable. 
From Eq. (\ref{consac}) the kinetic part of the second order action is 
\begin{equation}
\delta^{(2)} S_{\rm kin} = \int d^5x  \biggl\{ a^3\biggl[
- 6 \partial_{\alpha} \psi (\partial^{\alpha} \psi - \partial^{\alpha} \xi) 
+ \frac{1}{2} \partial_{\alpha} \chi \partial^{\alpha} \chi +[...]\biggr]
+[...]\biggr\}, 
\label{kinac}
\end{equation}
where the ellipses stand both for the other terms of (\ref{consac}) 
and for the five total derivatives. 
Eliminating now $\xi$ through Eq. (\ref{momcon}) 
we can see that Eq. (\ref{kinac}) can be written 
as 
\begin{eqnarray}
&& 
\delta^{(2)} S_{\rm kin} =  \int d^5 x\biggl\{ \frac{1}{2} \partial_{\alpha} 
{\cal G} \partial^{\alpha} {\cal G} + {\cal D}_{6} \biggr\},
\label{diakinac}\\
&& {\cal D}_{6} = \biggl[ - 3 \frac{a^3}{{\cal H}} \partial_{\alpha} \psi 
\partial^{\alpha} \psi \biggr]',
\nonumber
\end{eqnarray}
where ${\cal D}_{6}$ is a total derivative and ${\cal G}$ is given by 
\begin{equation}
{\cal G} = a^{3/2} \chi - z\psi,~~~~~z = \frac{a^{3/2} \varphi'}{{\cal H}}.
\label{canonical}
\end{equation}
Hence, ${\cal G}$ diagonalizes the kinetic 
part of the action  (\ref{consac}). It will now be shown that 
the same variable ${\cal G}$ diagonalizes the {\em full } action 
(\ref{consac}).

Notice, preliminary, that the variable ${\cal G}$ is gauge-invariant.
In fact, reading-off, from Eqs. (\ref{psil}) and (\ref{gvchi}), 
 the gauge-variations of $\psi$ and $\chi$ for infinitesimal 
diffeomorphisms we also see
that
\begin{equation}
\tilde{\cal G}\equiv {\cal G} = a^{3/2} X - z \Psi,
\end{equation}
where $\Psi$ and $X$ are, respectively,  the gauge-invariant
longitudinal fluctuation and the wall perturbation as defined in
Eqs. (\ref{giscal0}) and (\ref{chigi}). 

From Eqs. (\ref{momcon}) and (\ref{canonical}) 
the wall fluctuation and the derivative of the longitudinal 
fluctuation of the metric can be expressed as 
\begin{eqnarray}
&& \chi = \frac{ {\cal G}}{a^{3/2}} + \biggl( 
\frac{ \varphi'}{{\cal H}} \biggr)\psi,
\nonumber\\
&& \psi' = - {\cal H}\biggl[ \xi + \frac{{\varphi '}^2}{ 6 {\cal H}^2} \psi 
\biggr] - \frac{\varphi'}{6} \biggl( \frac{{\cal G}}{a^{3/2}} \biggr).
\label{trans1}
\end{eqnarray}
Inserting Eqs. (\ref{trans1}) into Eq. (\ref{consac}) 
we find, after a rather straightforward but algebraically long 
procedure, that 
\begin{eqnarray}
\delta^{(2)} S &=& \delta^{(2)}S_{\rm gr} + \delta^{(2)} S_{\rm m}
\nonumber\\
&=&  \int d^{5} x  \biggl\{ \frac{1}{2} \biggl[\partial_{\alpha} 
{\cal G} \partial^{\alpha} {\cal G} - {{\cal G}'}^2 - \frac{z''}{z} {\cal G}^2
\biggr]+ \sum_{i = 1}^{7} {\cal D}_{i} \biggr\},
\label{diagac}
\end{eqnarray}
where ${\cal D}_{7}$ is the last total derivative 
\begin{eqnarray}
{\cal D}_{7} &=& \biggl\{ \frac{{\varphi'}^2}{{\cal H}} a^3 \xi \psi - 
\frac{a^4}{2} \biggl( \frac{\varphi '}{{\cal H}} \biggr) 
 \biggl( \frac{\varphi '}{a} \biggr)' \psi^2  
\nonumber\\
&-& \varphi' a^{3/2} \xi {\cal G}-
\frac{a^{5/2}}{{\cal H}} \biggl( \frac{\varphi '}{a} \biggr)' {\cal G} \psi
- \frac{a^3}{6} \frac{ {\varphi'}^3}{{\cal H}^2} 
 \biggl(\frac{ \varphi''}{\varphi'} 
- {\cal H}\biggr) \psi\biggl(\frac{{\cal G}}{a^{3/2}}\biggr)\biggr\}'.
\end{eqnarray}

\renewcommand{\theequation}{5.\arabic{equation}}
\setcounter{equation}{0}
\section{Corrections to Newtonian potential}
The correction to Newton's law at short distances is 
well known in the case of the tensor modes. 
In the case of tensor fluctuations the second order 
action is well known and it is, for each polarization, 
\begin{equation}
\delta^{(2)} S_{(T)} = \int d^{5} x \biggl\{ \frac{1}{2} 
\biggl[\partial_{\alpha} 
\mu \partial^{\alpha} {\mu} - {{\mu}'}^2 - \frac{(a^{3/2})''}{a^{3/2}} \mu^2
\biggr]\biggr\},
\label{tm}
\end{equation}
where $\mu = \sqrt{2} h a^{3/2}$ and $h$ stands for each polarization of the 
tensor modes of the metric.
In general terms, the equation for the mass eigenstates
of the 
tensor normal modes can be written as 
\begin{equation}
\mu_{m}'' + \biggl[ m^2 - \frac{(a^{3/2})''}{a^{3/2}}\biggr] \mu_{m} =0,
\label{mass}
\end{equation}
and the related equation for $h_{m}$ is 
\begin{equation}
h_{m}'' + 3 {\cal H} h_{m}' + m^2 h_{m} =0.
\end{equation}

The tensor zero mode is always normalized since the integral
\begin{equation}
\int_{0}^{\infty} |\mu_{0}|^2 dw =\int_{0}^{\infty} dw a^{3}(w),
\end{equation}
is always convergent
provided the four-dimensional Planck mass  
\begin{equation}
M_{P}^2 = M^2 \int_{0}^{\infty} a^{3}(w) ~dw,
\end{equation}
is finite.
In the case of the higher-dimensional kink solution we can focus the 
attention on the case where, according to the example of Eqs. 
(\ref{s1})--(\ref{s3}) ,
$a(w) = ( b^2 w^2 + 1)^{-1/2}$. In this case the four-dimensional Planck 
mass is clearly finite.
In Eq. (\ref{mass})
\begin{equation}
\frac{(a^{3/2})''}{a^{3/2}} \sim \frac{15}{4~w^2}, ~~~~~b w \geq 1.
\label{large}
\end{equation}
The solution for the continuum modes will be 
\begin{eqnarray}
 \mu_{m}  = \sqrt{w} \biggl[ A J_{2}( m w) + B Y_{2}(m w)\biggr],
\label{lesol2}
\end{eqnarray}
where $J_{\nu}( mw)$ and $Y_{\nu}(m w)$ are Bessel functions of index $\nu$
\cite{abr}.
The solution given in Eq. (\ref{lesol2}) is the same one appearing 
in the case of Ref. \cite{3,3a} and 
determines the known correction to Newton's law
\begin{equation}
V(r) \sim G_{N} \frac{m_1 ~m_{2}}{r} \biggl[ 1 + \frac{1}{(b r)^2}\biggr],
\end{equation}
which arises from the contribution of the bulk continuum modes.

In the scalar case, the equation obeyed by the 
normal modes is given by 
\begin{equation}
{\cal G}_{m}'' + \biggl[ m^2 - \frac{z''}{z} \biggr] {\cal G}_{m} =0.
\label{G}
\end{equation}
As an example, consider the solution given in Eqs. (\ref{s1})--(\ref{s3}) where
\begin{equation}
z(w) = \frac{a^{3/2} \varphi'}{{\cal H}}= 
-\left( \frac{{\sqrt{6}}}{b\,w\,{\left( 1 + b^2\,w^2 \right) 
}^{\frac{3}{4}}} \right).
\label{zex}
\end{equation}
As previously discussed \cite{5}, 
the scalar zero mode is not normalizable. The solution 
for the zero mode is  
\begin{equation}
{\cal G}_{0}(w) = c_1 z(w) + c_2 z(w) \int^{w} \frac{dw'}{z^2(w')},
\end{equation}
and the integrand of
\begin{equation}
\int_{0}^{\infty} |{\cal G}_{0}|^2 dw, 
\end{equation}
diverges for $w\to 0$ for both linearly independent solutions 
parametrized by the two arbitrary constants $c_1$ and $c_2$. 
Notice also, incidentally, that $z''/z$ diverges when the zero-mode diverges..
This means that the zero mode is decoupled from the four dimensional dynamics.
If we want to discuss the corrections coming from the continuum modes 
it is useful to work with the field $g_{m} = (1/z) {\cal G}_{m}$ whose 
equation is 
\begin{equation}
g_{m}'' + 2 \frac{z'}{z} g_m' + m^2 g_{m} =0.
\label{g}
\end{equation}
The differential operator of Eq. (\ref{g})
is self-adjoint provided 
\begin{equation}
\left.\frac{d g_{m} }{dw}\right|_{1/b} =0,~~~~~~~~~
\left.\frac{d g_{m} }{dw}\right|_{w_{\rm max}} =0
\label{boundary}
\end{equation}
The effective cut-off $w_{\rm max}$ will be taken to $\infty$ after 
having determined 
the spectrum of mass eigenstates which is discrete for finite $w_{\rm max}$ 
but becomes continuous for $w_{\rm max}\to \infty$.

The solution for the massive modes can be obtained by noticing that 
\begin{equation}
\frac{z''}{z} \sim \frac{35}{4 w^2 }, ~~~~~w \geq 1/b.
\end{equation}
Consequently, from Eqs. (\ref{G}) and (\ref{g})   
\begin{eqnarray}
&&{\cal G}_{m}(w) = \sqrt{w} \biggl[ A_{S} J_{3}(m w) 
+ B_{S} Y_{3} (m w) \biggr], ~~~~b w \geq 1
\label{sB1}\\
&& g_{m}(w) =  \sqrt{w} (b w)^{5/2}  
\biggl[ A_{S} J_{3}(m w) + B_{S} Y_{3} (m w) \biggr], ~~~~b w \geq 1
\end{eqnarray}
Imposing now the boundary conditions (\ref{boundary})
we have that 
\begin{equation}
 A_{S} = - B_{S} \frac{Y_{2}(m/b)}{J_{2}(m/b)}.
\label{f1}
\end{equation}
At infinity the boundary conditions 
imply, instead,
\begin{equation}
A_S = - B_S \frac{Y_{2}(m w_{\rm max})}{J_{2}(m w_{\rm max})}.
\label{f2}
\end{equation}
Equating Eqs. (\ref{f1}) and (\ref{f2}) we find that 
\begin{equation}
\frac{Y_{2}(m/b)}{J_{2}(m/b)}  
= \frac{Y_{2}(m w_{\rm max})}{J_{2}(m w_{\rm max})}
\label{eq}
\end{equation}
which allows to determine the spectrum of mass eigenstates 
by noticing that at the right hand side  Eq. (\ref{eq})
the Bessel functions have a  {\em large} argument (i.e. $m w_{\rm max}$), 
whereas at the left hand side the Bessel functions have a {\rm small} 
argument, 
i.e. $m/b \ll 1$. Consequently, if the small and large argument limit 
is taken appropriately \cite{abr,tr} in Eq. (\ref{eq}), the resulting relation 
leads to the mass spectrum
\begin{equation}
m_{n} \simeq \frac{\pi}{2}( \frac{3}{2} + 2 n) \frac{1}{w_{\rm max}}
, ~~~~~~~n = 1, 2,3...
\end{equation}
which becomes continuous in the limit $w_{\rm max} \to \infty$.
The normalization condition 
\begin{equation}
\int_{1/b}^{\infty} {\cal G}_{m}(w) {\cal G}_{m'}(w) dw \equiv 
\int_{1/b}^{\infty} z^2(w) g_{m}(w) g_{m'}(w) dw = \delta(m-m')
\end{equation}
can be used in order to determine $B_{S}$.
the correction to the Newton's potential will be given by resumming 
\begin{equation}
V(r) \sim G_{N} \frac{m_1 ~m_{2}}{r} \biggl[ 1 + \frac{\pi}{2 b w_{\rm max}}
 \sum_{n} 
\biggl(\frac{m}{b}\biggr)^3 e^{- m_{n} r}\biggr].
\end{equation}
Transforming now the sum in an integral\footnote{Recall that $\delta m_{n} = 
m_{n + 1} - m_{n} = \pi/w_{\rm max}$. For $w_{\rm max} \to \infty$ 
 $(\pi/w_{\rm max}) 
\sum_{n} \to \int d m$. } and taking, consequently, the limit 
$w_{\rm max} \to \infty$ it is found that 
\begin{equation}
V(r) \sim G_{N} \frac{m_1 ~m_{2}}{r} \biggl[ 1 + 
\frac{3}{(b r)^4}\biggr].
\end{equation}
Hence, in the case of the specific example discussed in the present section, 
 the correction to the Newtonian potential coming from 
the bulk (scalar) continuum modes are more suppressed than the 
corrections coming from the tensor continuum modes.   This 
situation is reminiscent of what happens in the case of six-dimensional 
solutions when a string-like defect is included in the matter sector 
\cite{def,def1,def2,def3,gs}. 
The difference 
is that in the case of \cite{gs} the suppressed contribution 
comes from the tensor modes (in six dimensions), whereas,
in the present five-dimensional context, it comes from the 
scalar modes.

\renewcommand{\theequation}{6.\arabic{equation}}
\setcounter{equation}{0}
\section{Concluding remarks}
In this paper the scalar effective action for the normal modes 
of five-dimensional kink solutions has been derived. By perturbing 
the full action to second order in the amplitude 
of the scalar fluctuations the action for the scalar modes 
is 
\begin{equation}
\delta^{(2)} S =  \int d^{5} x  \biggl\{ \frac{1}{2} \biggl[\partial_{\alpha} 
{\cal G} \partial^{\alpha} {\cal G} - {{\cal G}'}^2 
- \frac{z''}{z} {\cal G}^2
\biggr]\biggr\}.
\end{equation}
The action is then expressed in terms of a single gauge-invariant fluctuation
\begin{eqnarray}
&&{\cal G}(x^{\mu},w) = a^{3/2}(w) X(x^{\mu},w) - z(w) \Psi(x^{\mu},w),
\nonumber\\
&& z(w) = \frac{a^{3/2} \varphi'}{{\cal H}},
\label{givar}
\end{eqnarray}
where $X$ is the gauge-invariant wall fluctuation and $\Psi$ is 
the gauge-invariant longitudinal fluctuation of the metric. Furthermore 
$a(w)$ is the warp factor and $\varphi$ is the kink background.
 
It is interesting to notice that neither the longitudinal 
fluctuations of the metric nor the wall fluctuations are the correct 
normal modes. The correct normal mode is obtained 
through a combination (with background-dependent coefficients) of the wall 
fluctuation and of the longitudinal metric perturbation.
The variable (\ref{givar}) is independent 
on the specific choice of coordinate system since it is 
invariant under infinitesimal diffeomorphisms. 
Furthermore, it should be appreciated that 
the derivation of (\ref{givar}) only assumes the validity of the background 
equations of motion and not of any {\em specific} 
background solution.
The variable (\ref{givar}) is the correct quantity 
to use in order to discuss the possible effects of scalar fluctuations 
in different frameworks. The zero mode associated with ${\cal G}$ is 
not localized but, still, the massive modes can lead to
corrections to Newton's law at short distances which have been computed 
in the case of a specific kink configuration. 
\section*{Acknowledgments}
It is a pleasure to acknowledge M. Shaposhnikov for useful discussions.

\newpage
\begin{appendix}
\renewcommand{\theequation}{A.\arabic{equation}}
\setcounter{equation}{0}
\section{Second order fluctuations of the geometry} 
In this Appendix the first and second order fluctuations
of the inverse metric and of the Christoffel connections will be reported.
They are  needed in order to obtain the 
second order form of the gravity and matter part of the action.

The scalar fluctuations of the metric can be written as 
\begin{equation}
\delta^{(1)} G_{A B}=a^2(w) \left(\matrix{2 ( \eta_{\mu\nu} \psi 
+ 2 \partial_{\mu} \partial_{\nu} E
& \partial_{\mu} C &\cr
\partial_{\mu} C  & 2 \xi &\cr}\right).
\end{equation}
The first order fluctuations of the inverse metric are 
\begin{eqnarray}
&&\delta^{(1)} G^{\mu\nu} = - \frac{2}{a^2} (\eta^{\mu\nu} \psi 
+ \partial^{\mu\nu} E), 
\nonumber\\
&& \delta^{(1)} G^{\mu\ w} = \frac{\partial^{\mu}C}{a^2}, 
\nonumber\\
&& \delta^{(1)} G^{w w} = - \frac{2 \xi}{a^2}. 
\label{m1}
\end{eqnarray}
The first order fluctuation 
of square root of the the determinant of the metric are
\begin{equation}
\delta^{(1)} \sqrt{|G|} = a^5 \biggl[ \Box E + 4 \psi- \xi
\biggr].
\label{d1}
\end{equation}
In order to perturb consistently the action (\ref{action}) 
the second order fluctuations of the inverse metric and of the 
square root of the determinant are needed. They are:
\begin{eqnarray}
&&\delta^{(2)} G^{\mu\nu} = \frac{4}{a^2} [  \psi^2 \eta^2 
+  \partial^{\mu}\partial^{\alpha} E 
+ 2 \psi \partial^{\mu}\partial^{\nu} E
\partial_{\alpha} \partial^{\nu} E]  - 
\frac{\partial^{\mu}C\partial^{\nu}C}{a^2},
\nonumber\\
&& \delta^{(2)} G^{\mu w} =  \frac{2}{a^2} \xi \partial^{\mu}C - 
\frac{2}{a^2} [ \eta^{\mu\alpha} \psi + \partial^{\mu\alpha} E] 
\partial_{\alpha}C,
\nonumber\\
&& \delta^{(2)} G^{w w} = \frac{1}{a^2} \partial_{\alpha}C 
\partial^{\alpha}C - 
\frac{ 4 }{a^2 } \xi^2.
\label{m2}
\end{eqnarray}
and 
\begin{equation}
\delta^{(2)} \sqrt{|G|} = \frac{a^5}{2} \biggl[ 8 \psi^2 + 4 \Box E \psi 
- \xi^2 - 8 \psi\xi + 
\partial_{\alpha} \partial^{\alpha} C - 2 \xi \Box E + (\Box E)^2 + 
\partial_{\alpha} \partial_{\beta} E \partial^{\alpha} \partial^{\beta} 
E \biggr].
\label{d2}
\end{equation}
The first order fluctuations of the Christoffel connections are 
\begin{eqnarray}
&& \delta^{(1)} \Gamma_{\mu\nu}^{w}= \eta_{\mu\nu}[ \psi' + 2 {\cal H}( \xi 
+ \psi)]  
 + \partial_{\mu}\partial_{\nu}[  E' + 2 {\cal H} E - C],
\nonumber\\
&& \delta^{(1)} \Gamma^{w}_{\mu w} =  \partial_{\mu}[ {\cal H} C - \xi ],
\nonumber\\
&& \delta^{(1)} \Gamma^{\mu}_{w w} = \partial^{\mu}[ C' + {\cal H} C -  
\xi ],
\nonumber\\
&& \delta^{(1)} \Gamma^{w}_{w w} = - \xi',
\nonumber\\
&& \delta^{(1)} \Gamma^{\mu}_{\alpha\beta} = 
 \partial_{\alpha}\psi \delta_{\beta}^{\mu} + 2\partial_{\beta} \psi 
\delta_{\alpha}^{\mu} -
\partial^{\mu}\psi \delta_{\alpha\beta} - {\cal H} \partial^{\mu}C 
\eta_{\alpha\beta} +
 \partial^{\mu}\partial_{\alpha} \partial_{\beta} E,
\nonumber\\
&& \delta^{(1)} \Gamma^{\alpha}_{\mu w} = \psi'\delta_{\mu}^{\alpha} + 
\frac{1}{2} \biggl( \partial_{\mu} \partial^{\alpha} C
- \partial^{\alpha} \partial_{\mu}C \biggr) +  
\partial_{\alpha}\partial^{\mu} E.
\label{c1}
\end{eqnarray}

Finally, the second order fluctuations of the Christoffel connections are 
\begin{eqnarray}
\delta^{(2)} \Gamma^{w}_{\mu \nu} &=& {\cal H} 
( 4 \xi^2 - \partial_{\alpha}C \partial^{\alpha}C ) \eta_{\mu\nu} 
+ \partial^{\alpha}C[ \partial_{\mu} 
H_{\nu\alpha} + \partial_{\nu} H_{\mu\alpha} - \partial_{\alpha} H_{\mu\nu} ]
- 2\xi  \partial_{\mu} \partial_{\nu} C 
\nonumber\\
 &+& 
2 \xi [ H_{\mu\nu}' + 2 {\cal H} H_{\mu\nu}],
\nonumber\\
\delta^{(2)} \Gamma^{w}_{w w} &=& {\cal H} \partial_{\alpha}C \partial^{\alpha}C + 
\partial_{\alpha}C {\partial^{\alpha} C'} - \partial_{\alpha}C \partial^{\alpha} \xi - 2 \xi \xi' ,
\nonumber\\
\delta^{(2)} \Gamma^{w}_{\mu w} &=& 2 {\cal H} \xi \partial_{\mu}C - 2 
\xi \partial_{\mu} \xi  +  \partial^{\alpha}C H_{\alpha\mu}' ,
\nonumber\\
\delta^{(2)} \Gamma^{\mu}_{w w} &=& \partial^{\mu}C \xi' - 2 {\cal H} 
H^{\mu \alpha} \partial_{\alpha}C  - 2 H^{\alpha\mu} \partial_{\alpha}C' + 2 H^{\mu\alpha} 
\partial_{\alpha} \xi,
\nonumber\\ 
\delta^{(2)} \Gamma^{\mu}_{ w \alpha} &=& -{\cal H}  \partial^{\mu}C 
\partial_{\alpha}C - 4 {\cal H} H^{\mu\beta} H_{\alpha\beta} 
- 2 H^{\mu\beta} H_{\alpha\beta}' 
+ \partial^{\mu}C \partial_{\alpha} \xi,
\nonumber\\
\delta^{(2)} \Gamma^{\mu}_{\alpha \beta} &=&  B^{\mu} 
\partial_{\beta} \partial_{\alpha} C  + 
2 {\cal H} [ H^{\mu\lambda}  \partial_{\lambda}C - \xi \partial^{\mu}C] \eta_{\alpha\beta} 
- H_{\alpha\beta}' \partial^{\mu}C - 2 {\cal H} \partial^{\mu}C H_{\alpha\beta} 
\nonumber\\
&+& 2 H^{\mu\lambda} ( \partial_{\lambda} H_{\alpha\beta} -
\partial_{\beta} H_{\lambda \alpha} - \partial_{\alpha} H_{\beta\lambda} ).
\label{c2}
\end{eqnarray}
where, in order to  reduce  the already lengthy expressions, the following 
quantity
\begin{equation}
H_{\mu\nu} = \psi \eta_{\mu\nu} + \partial_{\mu}\partial_{\nu} E,
\end{equation}
has been defined.
\end{appendix}

\end{document}